\documentclass[letterpaper]{article} 
\usepackage[preprint]{aaai2027}
\usepackage[hyphens]{url}  
\usepackage{graphicx} 
\usepackage{natbib}  
\usepackage{amsfonts} 
\usepackage{caption} 
\usepackage{algorithm}
\usepackage{algorithmic}
\usepackage{booktabs}
\usepackage{multirow}
\usepackage{newfloat}
\usepackage{listings}
\DeclareCaptionStyle{ruled}{labelfont=normalfont,labelsep=colon,strut=off} 
\floatstyle{ruled}
\newfloat{listing}{tb}{lst}{}
\floatname{listing}{Listing}

\usepackage{booktabs}

\title{What Makes Codec Tokens Effective for Target Speaker Extraction? A Semantic-Acoustic Disentanglement Perspective}

\author{
    Wang Xiang\textsuperscript{\rm 1},
    Bang Zeng\textsuperscript{\rm 1},
    Ming Li\textsuperscript{\rm 2}\corresponding
}
\affiliations{
    \textsuperscript{\rm 1}Wuhan University\\
    \textsuperscript{\rm 2}The Chinese University of Hong Kong, Shenzhen\\
    xiangw@whu.edu.cn, bangzeng@whu.edu.cn, mingli369@cuhk.edu.cn
}

\begin{document}

\maketitle

\begin{abstract}

 Target speaker extraction (TSE) aims to recover a target speaker's speech from a speech mixture using an enrollment utterance. Recent codec-based generative approaches formulate TSE as conditional speech token generation and achieve promising perceptual quality. However, most existing systems rely on unified codec tokens that entangle linguistic content, speaker characteristics, and acoustic details, forcing a single model to jointly handle target selection, content semantic, and acoustic reconstruction. In this paper, we investigate whether disentangled semantic–acoustic codec representations are better suited to TSE. We first show that simply replacing unified codec tokens with semantic-acoustic tokens is insufficient and can even degrade performance when both remain handled by a unified generative model. Motivated by this observation, we propose a task-aware discriminative–generative framework, where a discriminative branch extracts target-related semantic information, while a generative branch predicts coarse acoustic tokens and refines them under semantic guidance.  This specialization separates what is said from how it sounds while coordinating both factors in a factorized codec space. Experiments on Libri2Mix show that our method reduces dWER from 0.388 to 0.118 compared with a unified-codec baseline, while maintaining competitive perceptual quality and speaker consistency. Ablation studies confirm that the gains arise from task-aware semantic–acoustic modeling rather than representation disentanglement alone.

\end{abstract}


\section{Introduction}

\begin{figure}[t]
\centering
\includegraphics[width=\columnwidth]{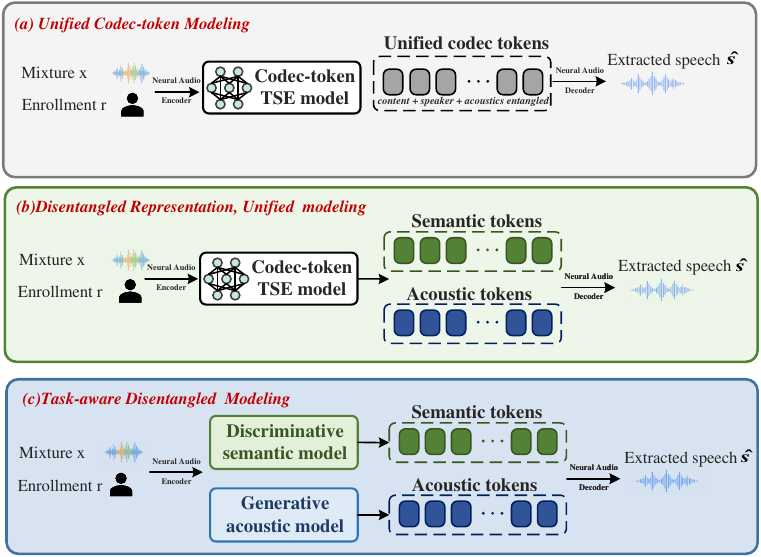}
\caption{
\textbf{Motivation for task-aware semantic--acoustic codec-token
modeling.}
(a) Unified codec-token modeling entangles linguistic content,
speaker characteristics, and acoustic details within a single
generative model.
(b) Representation-level disentanglement alone remains insufficient
when semantic and acoustic token streams are still handled by a
unified modeling paradigm.
(c) Our task-aware framework assigns discriminative modeling to target
semantics and generative modeling to acoustic reconstruction, aligning
each token stream with its distinct role in TSE.
}
\label{fig:intro_motivation}
\end{figure}

Target speaker extraction (TSE) aims to recover the speech of a target speaker from a multi-speaker mixture with the assistance of an enrollment utterance \cite{voicefilter,xu2020spex,zmolikova2023neural,spextse}. Unlike conventional
speech separation methods\cite{10856367}, which estimate all sources in a mixture \cite{convtasnet,dprnn,sepformer}, TSE only extracts the speaker specified by the enrollment speech, making it particularly suitable for human-centric applications such as hearing assistance, virtual assistant, and multi-party spoken interaction. 

Most existing TSE systems follow a discriminative paradigm, directly estimating the target waveform, spectrum, or time-frequency mask from the mixture and enrollment speech
\cite{voicefilter,spextse,wesep,usef,he2024hierarchical,wang2018multichannel,chen2017deep,wang2018supervised}. Benefiting from strong
target-conditioned feature extraction, these methods are generally effective at interference suppression and target-speaker control\cite{wang2024dualchannel,delcroix2020speakerbeam,hao2020unified}. However, they are commonly optimized with deterministic signal-level objectives, which may not fully capture the perceptual distribution of natural speech. Consequently, even when interference is successfully removed, the extracted speech may remain distorted, over-smoothed, or
limited in fine-grained acoustic details.

Generative TSE provides a complementary solution by modeling the distribution of target speech rather than a single deterministic estimate\cite{richter2023speech,lu2022conditional}. Recent approaches have explored diffusion models, variational models, and codec-token language models for speech enhancement and extraction
\cite{diffusionse,segan,tselm,anyenhance,speechx}. In particular, codec-based systems formulate TSE as conditional sequence generation: the target speech is encoded into discrete neural codec tokens, predicted by a language model, and reconstructed by a codec decoder\cite{lauratse}. These results establish codec tokens as a promising interface between target-conditioned speech processing and generative sequence modeling.

Nevertheless, existing codec-token TSE systems largely adopt unified
codec representations, in which linguistic content, speaker
characteristics, prosody, and acoustic details are encoded in the same
token space, forcing a single model to simultaneously handle target selection, content preservation, speaker consistency, and acoustic reconstruction. A natural alternative is to use a disentangled codec that
separates semantic and acoustic representations. However,
representation-level disentanglement does not necessarily yield
modeling-level disentanglement. If semantic and acoustic streams are
still jointly handled by a single generative model, their distinct roles
are effectively coupled again during prediction. Indeed, our controlled
comparison shows that directly replacing the unified codec in a
generative TSE model with disentangled codec tokens does not
consistently improve performance. This observation suggests that the
effectiveness of codec tokens depends not only on how speech is
represented, but also on whether the modeling strategy is aligned with
the structure of that representation.

This distinction is particularly important for TSE because the task contains two complementary but fundamentally different objectives. First, the system must determine \emph{what is said} by the target speaker, requiring robust target selection, interference rejection, and linguistic-content preservation. Errors in this aspect lead to word deletions, substitutions, hallucinations, or leakage from interfering
speakers. Second, the system must reconstruct \emph{how it sounds}, including naturalness, timbre, prosody, and fine-grained acoustic details. These two objectives impose different modeling requirements: target selection and content preservation require reliable target-conditioned estimation, whereas acoustic reconstruction involves modeling speaker characteristics and fine-grained speech details. Handling both streams with a single generative model couples their distinct roles during prediction and may prevent the model from fully exploiting the semantic–acoustic structure.


Motivated by this analysis, we propose a task-aware semantic–acoustic discriminative–generative framework for TSE. The target speech is represented by separate semantic and acoustic tokens. A discriminative extractor first estimates the target speech, from which a frozen semantic tokenizer derives a target-related semantic representation for target selection and content preservation. In parallel, a decoder-only language model predicts coarse acoustic tokens from the mixture and enrollment representations. A semantic-conditioned refiner then uses the coarse acoustic tokens and the semantic representation to recover the complete acoustic embedding. The semantic and refined acoustic embeddings are combined in the codec space and decoded to reconstruct the target waveform. our approach coordinates the two paradigms directly in a factorized codec space. Figure~\ref{fig:intro_motivation} summarizes the distinction among unified codec-token modeling, disentangled representation but unified modeling, and the proposed task-aware disentangled modeling.

Experiments demonstrate that the proposed framework
achieves a stronger balance among perceptual quality, intelligibility,and speaker consistency than unified-codec generation and naive unified modeling of disentangled token. Ablation studies further show that the improvement does not arise from codec replacement alone, but from matching semantic and acoustic representations with suitable modeling paradigms. We additionally analyze how split and joint optimization, as well as autoregressive and non-autoregressive
inference, affect the collaboration between the two tokens.

Our main contributions are summarized as follows:

\begin{itemize}
    \item We investigate what makes codec tokens effective for
    target speaker extraction from a semantic--acoustic disentanglement
    perspective. Our study reveals that representation-level
    disentanglement alone is insufficient: directly replacing unified codec with factorized semantic and acoustic
    tokens may even degrade TSE performance when the two streams remain
    handled by a unified generative model. Effective codec-token TSE
   requires not only structured representations, but also a
    modeling strategy aligned with their distinct roles.

    \item Guided by this finding, we propose a task-aware
    discriminative--generative framework for semantic--acoustic codec
    modeling. A discriminative branch extracts target-related semantic
    information for reliable speaker selection and content preservation,
    while a generative branch predicts coarse acoustic tokens and
    refines them under semantic guidance to recover naturalness,
    speaker characteristics, and fine-grained acoustic details.

    \item We design controlled comparisons among unified-codec
    modeling, unified generative modeling of disentangled
    token, and task-aware disentangled modeling. The results
    validate our central finding: the effectiveness of codec
    tokenization for TSE hinges on matching its representation structure
 with appropriate task-specific modeling, rather than relying on
    semantic--acoustic disentanglement at the representation-level alone.
\end{itemize}

\section{Related Work}

\subsection{Target Speaker Extraction}

Most target speaker extraction systems follow a discriminative
paradigm, directly estimating the target waveform, spectrum, or
time--frequency mask from the mixture and enrollment speech.
Early methods employ speaker embeddings extracted from the enrollment
utterance to condition the extraction network
\cite{voicefilter,spextse,xiang26_jsjtu,10856367}. Subsequent advances in time-domain and
time--frequency modeling have substantially improved interference
suppression and target-speaker preservation
\cite{convtasnet,dprnn,sepformer,wesep}. More recently,
speaker-embedding-free approaches directly model the interaction
between mixture and enrollment representations, reducing the mismatch
between speaker verification and target extraction
\cite{usef}.

Generative approaches instead model the distribution of clean speech
to improve perceptual quality and restore missing acoustic details.
Recent studies have explored diffusion models and discrete-token
language models for speech enhancement and target speaker extraction
\cite{diffusionse,tselm,anyenhance}. In particular, codec-based
approaches formulate TSE as conditional speech-token generation, using
language models to predict target codec representations
\cite{lauratse}. Discriminative--generative systems further combine
the target controllability of discriminative extraction with the
reconstruction capability of generative models
\cite{usef_lauratse}. However, existing approaches either generate
unified codec representations or combine the two paradigms without
explicitly assigning semantic and acoustic factors to specialized
token-level models.

\subsection{Neural Audio Codecs for Speech Modeling}

Neural audio codecs convert continuous speech into discrete
representations through vector quantization or residual vector
quantization. Representative systems, including SoundStream\cite{soundstream}, EnCodec\cite{encodec},
and FunCodec\cite{funcodec}, encode speech into one or more codebook streams and
reconstruct the waveform with a neural decoder. Their discrete and compact
representations have enabled language-model-based speech generation,
enhancement, and target speaker extraction. Nevertheless, these codecs
are primarily optimized for reconstruction fidelity, and their tokens
generally encode linguistic content, speaker characteristics, prosody,
and acoustic details in a unified representation.

Recent codecs introduce more structured representations by separating
different speech factors. DualCodec, for example, decomposes speech
into semantic and acoustic token streams, providing content-related
representations and complementary acoustic details
\cite{dualcodec}. While such codecs demonstrate the benefits of
representation-level disentanglement for speech generation, how their
factorized token spaces should be modeled for target speaker extraction
remains underexplored. 

\begin{figure*}
    \centering
    \includegraphics[width=0.83\textwidth]{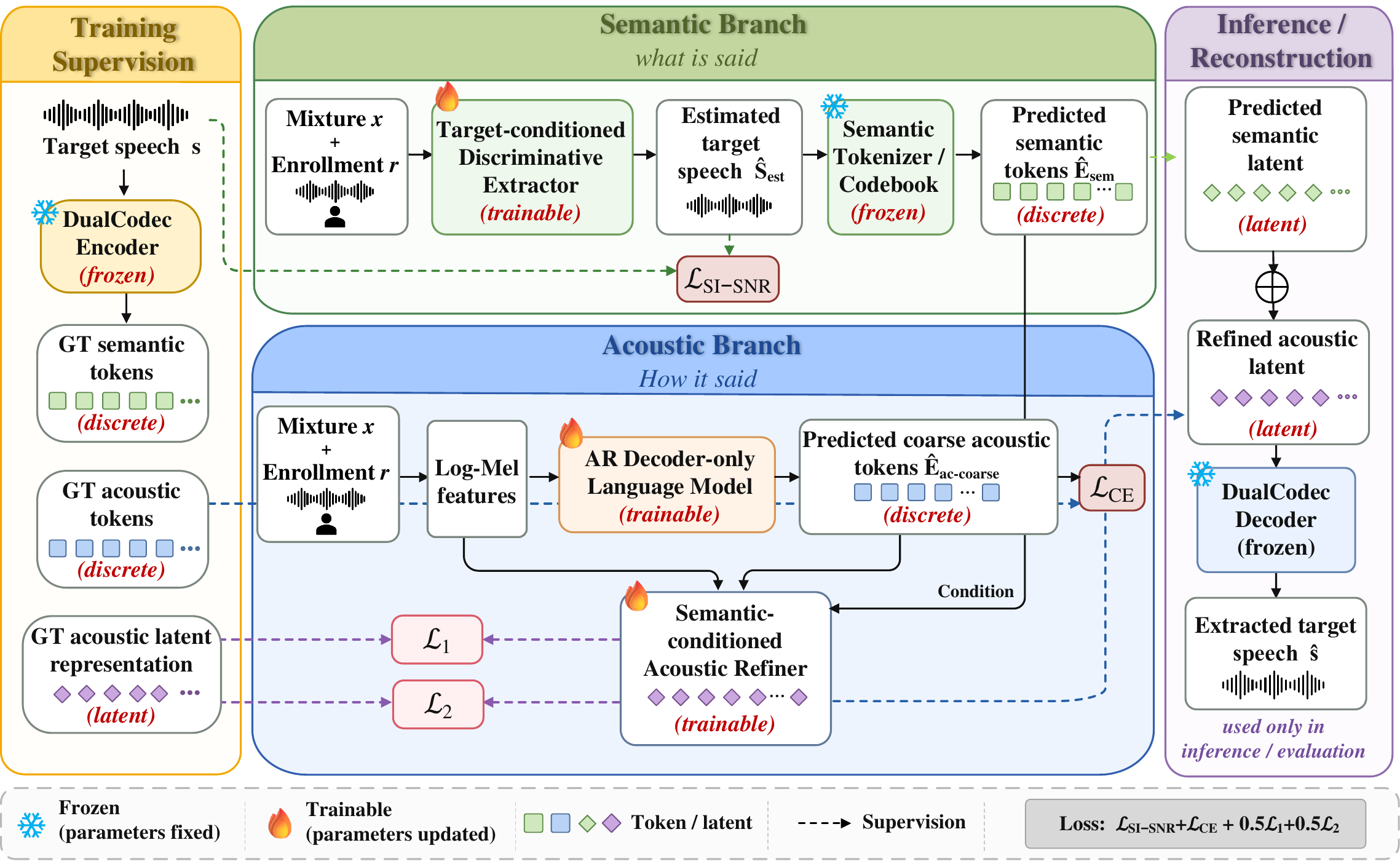}
    \caption{
\textbf{Overview of the proposed framework.} A discriminative branch provides target-related semantic representations, while an autoregressive model predicts coarse acoustic tokens. The semantic-conditioned refiner recovers the complete acoustic embedding, which is combined with the semantic representation for waveform reconstruction.}
    \label{fig:双栏大图}
\end{figure*}

\section{Method}

\subsection{Overview}

Given a mixture speech signal
$x\in\mathbb{R}^{T_x}$ and an enrollment utterance
$r\in\mathbb{R}^{T_r}$ from the target speaker, target speaker
extraction aims to recover the clean target speech
$s\in\mathbb{R}^{T_s}$. Instead of modeling a unified codec, we represent the target speech using disentangled
semantic-acoustic tokens with Dualcodec\cite{dualcodec}.


As illustrated in Figure~\ref{fig:双栏大图}, the proposed framework consists of two task-specialized branches: a discriminative semantic branch and a generative acoustic branch. The semantic branch first estimates the target waveform from the mixture and enrollment utterance using a target-conditioned discriminative extractor. A frozen semantic tokenizer then converts the estimated waveform into a target-related semantic representation, providing reliable target selection and content preservation. In parallel, the acoustic branch uses a decoder-only language model to predict the first several acoustic codebook streams from the mixture and enrollment representations. These predicted tokens form a coarse acoustic representation, which is subsequently processed by a semantic-conditioned refiner to estimate the complete acoustic embedding. In this way, the semantic branch primarily preserves what is said by the target speaker, while the acoustic branch reconstructs how the target speech sounds.


During training, the discriminative extractor is supervised by SI-SNR, while the acoustic branch is optimized with token classification and embedding reconstruction losses. During inference, the target-related semantic embedding and refined acoustic embedding are combined and decoded into the target waveform.

\subsection{Semantic--Acoustic Codec Representation}

We employ a pretrained DualCodec to represent speech using two
complementary streams. Given the clean target speech $s$, the frozen
codec encoder produces semantic and acoustic token sequences:
\begin{equation}
    Z^{\mathrm{sem}},Z^{\mathrm{aco}}
    =
    Q_{\mathrm{dual}}(s),
    \label{eq:dualcodec_encoding}
\end{equation}
where
\begin{equation}
    Z^{\mathrm{sem}}
    \in
    \{1,\ldots,K_{\mathrm{sem}}\}^{T}
\end{equation}
denotes the content-dominant semantic-token sequence, and
\begin{equation}
    Z^{\mathrm{aco}}
    \in
    \{1,\ldots,K_{\mathrm{aco}}\}^{T\times Q}
\end{equation}
denotes the acoustic-token sequence with $Q$ residual quantization
layers. Here, $T$ is the codec-frame length, while
$K_{\mathrm{sem}}$ and $K_{\mathrm{aco}}$ denote the corresponding
codebook sizes.

The semantic stream mainly preserves content-related information,
whereas the acoustic stream provides complementary speaker, prosodic,
and fine-grained details required for high-quality reconstruction. The
complete ground-truth acoustic embedding is obtained through acoustic
codebook lookup:
\begin{equation}
    E^{\mathrm{aco}}
    =
    B_{\mathrm{aco}}
    \left(
        Z^{\mathrm{aco}}
    \right)
    \in
    \mathbb{R}^{T\times d_{\mathrm{aco}}},
    \label{eq:gt_acoustic_embedding}
\end{equation}
where $B_{\mathrm{aco}}(\cdot)$ denotes the frozen acoustic codebook
embedding operation.

During training, $Z^{\mathrm{aco}}$ and $E^{\mathrm{aco}}$ are used as
the supervision targets for acoustic-token generation and acoustic
refinement, respectively.

\subsection{Target-Conditioned Discriminative Extractor}

The discriminative branch is responsible for target-speaker
selection, interference suppression, and target-content preservation.
Given the mixture $x$ and enrollment utterance $r$, a
target-conditioned discriminative extractor estimates the target
waveform:
\begin{equation}
    \widetilde{s}
    =
    F_{\mathrm{dis}}(x,r),
    \label{eq:discriminative_estimation}
\end{equation}
where $F_{\mathrm{dis}}(\cdot)$ denotes the discriminative target
extraction model and $\widetilde{s}$ is the estimated target speech.

The discriminative branch is optimized using the negative
scale-invariant signal-to-noise ratio:
\begin{equation}
    \mathcal{L}_{\mathrm{SI\mbox{-}SNR}}
    =
    -
    \mathrm{SI\mbox{-}SNR}
    \left(
        \widetilde{s},s
    \right).
    \label{eq:sisnr_loss}
\end{equation}
This objective encourages the discriminative branch to preserve the
target signal while suppressing interfering speakers.

The estimated target speech is subsequently processed by the frozen
semantic encoder and codebook of DualCodec:
\begin{equation}
    \widehat{Z}^{\mathrm{sem}}
    =
    Q_{\mathrm{sem}}
    \left(
        \widetilde{s}
    \right),
    \label{eq:pred_semantic_tokens}
\end{equation}
\begin{equation}
    \widehat{E}^{\mathrm{sem}}
    =
    B_{\mathrm{sem}}
    \left(
        \widehat{Z}^{\mathrm{sem}}
    \right),
    \label{eq:pred_semantic_embedding}
\end{equation}
where $Q_{\mathrm{sem}}(\cdot)$ and
$B_{\mathrm{sem}}(\cdot)$ denote the frozen semantic tokenizer and
semantic codebook embedding operation, respectively.


\subsection{Autoregressive Acoustic Token Generation}

The generative branch predicts coarse acoustic codec tokens from the
mixture and enrollment speech. The two input waveforms are first
converted into log-Mel spectrograms and processed by a weight-shared
acoustic encoder:
\begin{equation}
    E_x
    =
    C\left(\mathrm{Mel}(x)\right),
    \qquad
    E_r
    =
    C\left(\mathrm{Mel}(r)\right),
    \label{eq:acoustic_encoder}
\end{equation}
where $C(\cdot)$ denotes the shared Conformer encoder.

To reduce the complexity of autoregressive generation, the
decoder-only language model predicts only the first $Q_p$ acoustic
quantization layers, where $Q_p\leq Q$:
\begin{equation}
    Z^{\mathrm{aco}}_{1:Q_p}
    \in
    \{1,\ldots,K_{\mathrm{aco}}\}^{T\times Q_p}.
\end{equation}
These acoustic token streams are serialized into a one-dimensional
sequence:
\begin{equation}
    Y
    =
    \mathcal{S}
    \left(
        Z^{\mathrm{aco}}_{1:Q_p}
    \right)
    =
    (y_1,\ldots,y_N),
    \label{eq:acoustic_serialization}
\end{equation}
where $\mathcal{S}(\cdot)$ denotes the token serialization operation
and $N$ is the resulting sequence length.

The acoustic-token distribution is factorized autoregressively as
\begin{equation}
    p_{\theta_{\mathrm{aco}}}
    \left(
        Y\mid E_x,E_r
    \right)
    =
    \prod_{n=1}^{N}
    p_{\theta_{\mathrm{aco}}}
    \left(
        y_n
        \mid
        y_{<n},E_x,E_r
    \right).
    \label{eq:ar_acoustic_distribution}
\end{equation}
Therefore, coarse acoustic-token generation is conditioned only on
the mixture representation, enrollment representation, and previously
generated acoustic tokens. The semantic representation is introduced
later in the acoustic refinement stage.

During training, teacher forcing is employed, and the acoustic-token
classification objective is
\begin{equation}
    \mathcal{L}_{\mathrm{CE}}
    =
    -
    \frac{1}{N}
    \sum_{n=1}^{N}
    \log
    p_{\theta_{\mathrm{aco}}}
    \left(
        y_n
        \mid
        y_{<n},E_x,E_r
    \right).
    \label{eq:acoustic_ce_loss}
\end{equation}

During inference, the acoustic tokens are generated autoregressively
and reorganized into their original multi-codebook structure:
\begin{equation}
    \widehat{Z}^{\mathrm{aco}}_{1:Q_p}
    =
    \mathcal{S}^{-1}
    \left(
        \widehat{Y}
    \right).
    \label{eq:pred_acoustic_tokens}
\end{equation}

\subsection{Semantic-Conditioned Acoustic Refinement}

The generated coarse acoustic tokens contain the primary acoustic
structure of the target speech, but predicting only a subset of the
quantization layers may lose fine-grained acoustic information. We
therefore introduce an encoder-only refinement module to recover the
complete acoustic embedding.

The predicted coarse acoustic tokens are first mapped to continuous
codebook embeddings:
\begin{equation}
    \widehat{E}^{\mathrm{aco}}_{\mathrm{coarse}}
    =
    B_{\mathrm{aco}}
    \left(
        \widehat{Z}^{\mathrm{aco}}_{1:Q_p}
    \right).
    \label{eq:coarse_acoustic_embedding}
\end{equation}
The refinement module then combines the coarse acoustic embedding with
the mixture representation, enrollment representation, and
target-related semantic condition:
\begin{equation}
    \widetilde{E}^{\mathrm{aco}}
    =
    R
    \left(
        E_x,
        E_r,
        \widehat{E}^{\mathrm{aco}}_{\mathrm{coarse}},
        \widehat{E}^{\mathrm{sem}}
    \right),
    \label{eq:acoustic_refinement}
\end{equation}
where $R(\cdot)$ denotes the encoder-only acoustic refiner and
\begin{equation}
    \widetilde{E}^{\mathrm{aco}}
    \in
    \mathbb{R}^{T\times d_{\mathrm{aco}}}
\end{equation}
is the refined acoustic embedding.

Unlike the autoregressive generator, the refinement module explicitly
uses the target-related semantic representation. This guidance helps
the model maintain target content and speaker consistency while
recovering acoustic details that are not captured by the generated
coarse token layers.

The refiner is supervised using the complete acoustic embedding
$E^{\mathrm{aco}}$ extracted from the clean target speech. Let
$m_t\in\{0,1\}$ denote the valid-frame mask for variable-length
utterances. The $L_1$ reconstruction loss is defined as
\begin{equation}
    \mathcal{L}_{1}
    =
    \frac{
        \sum_{t=1}^{T}
        m_t
        \left\|
            \widetilde{E}^{\mathrm{aco}}_t
            -
            E^{\mathrm{aco}}_t
        \right\|_1
    }{
        \sum_{t=1}^{T}m_t
    },
    \label{eq:l1_loss}
\end{equation}
and the $L_2$ reconstruction loss is defined as
\begin{equation}
    \mathcal{L}_{2}
    =
    \frac{
        \sum_{t=1}^{T}
        m_t
        \left\|
            \widetilde{E}^{\mathrm{aco}}_t
            -
            E^{\mathrm{aco}}_t
        \right\|_2^2
    }{
        \sum_{t=1}^{T}m_t
    }.
    \label{eq:l2_loss}
\end{equation}

The two embedding losses encourage the refined representation to
recover both the overall acoustic structure and fine-grained residual
details of the clean target speech.

\subsection{Training Objective}

The overall training objective consists of the discriminative
target-extraction loss, the acoustic-token classification loss, and
the acoustic-embedding reconstruction losses:
\begin{equation}
    \mathcal{L}
    =
    \lambda_{\mathrm{SI\mbox{-}SNR}}
    \mathcal{L}_{\mathrm{SI\mbox{-}SNR}}
    +
    \lambda_{\mathrm{CE}}
    \mathcal{L}_{\mathrm{CE}}
    +
    \lambda_{1}
    \mathcal{L}_{1}
    +
    \lambda_{2}
    \mathcal{L}_{2},
    \label{eq:overall_loss}
\end{equation}
where
$\lambda_{\mathrm{SI\mbox{-}SNR}}$,
$\lambda_{\mathrm{CE}}$,
$\lambda_{1}$, and $\lambda_{2}$
control the contributions of the corresponding objectives.

The SI-SNR objective supervises the discriminative branch for target
extraction and interference suppression. The cross-entropy objective
supervises autoregressive acoustic-token generation, while the $L_1$
and $L_2$ objectives supervise semantic-conditioned acoustic
refinement.


\subsection{Inference and Waveform Reconstruction}

During inference, only the mixture speech $x$ and enrollment utterance
$r$ are required. First, the discriminative branch estimates the
target speech:
\begin{equation}
    \widetilde{s}
    =
    F_{\mathrm{dis}}(x,r).
\end{equation}
The frozen semantic tokenizer then extracts the target-related semantic
embedding:
\begin{equation}
    \widehat{E}^{\mathrm{sem}}
    =
    B_{\mathrm{sem}}
    \left(
        Q_{\mathrm{sem}}
        \left(
            \widetilde{s}
        \right)
    \right).
    \label{eq:inference_semantic}
\end{equation}

In parallel, the autoregressive language model predicts the coarse
acoustic token streams:
\begin{equation}
    \widehat{Z}^{\mathrm{aco}}_{1:Q_p}
    =
    F_{\mathrm{AR}}(x,r).
    \label{eq:inference_acoustic}
\end{equation}
The predicted coarse acoustic embedding is then refined under semantic
guidance:
\begin{equation}
    \widetilde{E}^{\mathrm{aco}}
    =
    R
    \left(
        E_x,
        E_r,
        B_{\mathrm{aco}}
        \left(
            \widehat{Z}^{\mathrm{aco}}_{1:Q_p}
        \right),
        \widehat{E}^{\mathrm{sem}}
    \right).
    \label{eq:inference_refinement}
\end{equation}

Finally, the semantic and acoustic representations are combined in the
DualCodec latent space:
\begin{equation}
    \widehat{H}
    =
    \widehat{E}^{\mathrm{sem}}
    +
    \widetilde{E}^{\mathrm{aco}},
    \label{eq:latent_combination}
\end{equation}
and the frozen DualCodec decoder reconstructs the extracted target
waveform:
\begin{equation}
    \widehat{s}
    =
    D_{\mathrm{dual}}
    \left(
        \widehat{H}
    \right).
    \label{eq:waveform_decoding}
\end{equation}

\begin{table*}[t]
\centering
\small
\setlength{\tabcolsep}{3.2pt}
\renewcommand{\arraystretch}{1.08}
\begin{tabular}{lcccccccccc}
\toprule
\multirow{2}{*}{Model}
& \multirow{2}{*}{Category}
& \multirow{2}{*}{Codecs}
& \multicolumn{3}{c}{DNSMOS $\uparrow$}
& \multirow{2}{*}{NISQA $\uparrow$}
& \multirow{2}{*}{SBERT $\uparrow$}
& \multirow{2}{*}{dWER $\downarrow$}
& \multirow{2}{*}{WavLM $\uparrow$}
& \multirow{2}{*}{WeSpeaker $\uparrow$} \\
\cmidrule(lr){4-6}
& & & SIG & BAK & OVRL & & & & & \\
\midrule

Mixture
& --
& --
& 3.383 & 3.098 & 2.653
& 2.453 & 0.572 & 0.792
& 0.847 & 0.759 \\

\midrule

X-TF-GridNet
& D
& --
& 3.328 & 3.607 & 2.895
& 3.216 & 0.784 & 0.389
& -- & -- \\

USEF-SepFormer
& D
& --
& 3.324 & 3.698 & 2.927
& 2.880 & 0.830 & 0.156
& -- & -- \\

SpEx+
& D
& --
& 3.472 & 4.027 & 3.186
& 3.349 & 0.878 & 0.148
& 0.973 & 0.935 \\

WeSep
& D
& --
& 3.486 & 3.838 & 3.118
& 3.892 & 0.895 & 0.123
& 0.980 & 0.945 \\

USEF-TFGridNet-L
& D
& --
& 3.555 & 4.051 & 3.272
& 4.319 & \textbf{0.935} & \textbf{0.0747}
& \textbf{0.988} & \textbf{0.968} \\

\midrule

TSELM-L
& G
& WavLM
& 3.489 & 4.041 & 3.212
& 3.961 & 0.793 & 0.297
& 0.887 & 0.627 \\

AnyEnhance
& G
& DAC
& 3.638 & 4.066 & 3.353
& 4.277 & 0.735 & --
& 0.914 & -- \\

LauraTSE
& G
& FunCodec
& 3.609 & 4.084 & 3.336
& \textbf{4.333} & 0.908 & 0.159
& 0.974 & 0.876 \\

LLaSE-G1
& G
& X-Codec2
& 3.531 & 4.015 & 3.226
& 3.638 & 0.825 & 0.476
& -- & -- \\

Metis
& G
& DAC-style
& 3.588 & 3.980 & 3.265
& 3.869 & 0.890 & 0.180
& -- & -- \\

GenTSE
& G
& SimCodec
& \textbf{3.656} & \textbf{4.135} & \textbf{3.399}
& 3.976 & 0.920 & 0.177
& -- & -- \\

\midrule

Ours-Unified
& G
& DualCodec
& 3.427 & 3.914 & 3.105
& 3.852 & 0.786 & 0.388
& 0.928 & 0.778 \\

\textbf{Ours-Specialized}
& D--G
& DualCodec
& 3.549 & 4.093 & 3.260
& 4.052 & 0.902 & 0.118
& 0.960 & 0.837 \\

\bottomrule
\end{tabular}
\caption{
Results on Libri2Mix clean. ``D'' and ``G'' denote discriminative
and generative models, respectively. ``Codecs'' denotes the neural audio codecs used by each generative system.
}
\label{tab:main_results}
\end{table*}

\subsection{Experimental Setup}

\paragraph{Datasets.}
 We use the 460-hour clean subset of LibriSpeech, denoted as LibriSpeech-460h, for large-scale pre-training \cite{panayotov2015librispeech}. Training mixtures are generated online by randomly selecting target and interfering utterances from different speakers and mixing them at a relative signal-to-noise ratio uniformly sampled between 0 and 5 dB. An additional utterance from the target speaker is used as the enrollment speech. We use the clean subsets of Libri2Mix for and evaluation. 

\paragraph{Baselines.}
We compare with representative discriminative TSE systems, including
SpEx+\cite{spextse}, WeSep\cite{wesep}, X-TF-GridNet\cite{hao2024x} and USEF-TSE\cite{usef}
; generative systems, including TSELM\cite{tselm},
AnyEnhance\cite{anyenhance}, LauraTSE\cite{lauratse}, LLaSE-G1\cite{kang2025llase}, Metis\cite{wang2026metis} and GenTSE\cite{li26_interspeech}.


\paragraph{Evaluation Metrics.}
We evaluate the systems from four complementary perspectives.
Intelligibility is measured by differential word error rate
(dWER, $\downarrow$), computed using
Whisper~\cite{radford2023robust}. Perceptual quality is assessed using
DNSMOS~\cite{reddy2022dnsmos} and
NISQA~\cite{mittag2021nisqa}. We use
SpeechBERTScore~\cite{saeki2024speechbertscore} to measure semantic
similarity, and cosine similarities based on
WavLM~\cite{chen2022wavlm} and
WeSpeaker~\cite{wang2023wespeaker} embeddings to evaluate speaker
consistency. Higher values indicate better performance for all metrics
except dWER.

\paragraph{Implementation Details.}
We adopt a two-stage training procedure. First, the discriminative semantic branch is pretrained on the Libri2Mix clean training set. It is then integrated with the generative acoustic branch using either split or joint training. Under split training, the pretrained semantic branch is frozen, and only the acoustic branch is optimized using token classification and embedding reconstruction losses. Under joint training, the complete framework is fine-tuned end-to-end. The semantic and acoustic branches contain approximately 16M and 77M parameters, respectively, including a 36M decoder-only language model in the acoustic branch. All models are trained for up to 100 epochs using Adam and trained on NVIDIA A800 80 GB GPUs.

\subsection{Main Results}

\begin{table*}[t]
\centering
\small
\setlength{\tabcolsep}{2.5pt}
\renewcommand{\arraystretch}{1.08}
\begin{tabular}{@{}lccccccccccc@{}}
\toprule
\multirow{2}{*}{Methods}
& \multirow{2}{*}{Codec}
& \multirow{2}{*}{Modeling}
& \multirow{2}{*}{Role}
& \multicolumn{3}{c}{DNSMOS $\uparrow$}
& \multirow{2}{*}{NISQA $\uparrow$}
& \multirow{2}{*}{SBERT $\uparrow$}
& \multirow{2}{*}{dWER $\downarrow$}
& \multirow{2}{*}{WavLM $\uparrow$}
& \multirow{2}{*}{WeSpk $\uparrow$} \\
\cmidrule(lr){5-7}
& & & & SIG & BAK & OVRL & & & & & \\
\midrule

\multicolumn{12}{l}{
\textit{(a) Oracle reconstruction of clean target speech}
} \\

\multirow{2}{*}{Oracle}
& FunCodec
& --
& --
& 3.534
& 3.894
& 3.185
& 4.164
& 0.981
& 0.080
& 0.991
& 0.930 \\

& DualCodec
& --
& --
& 3.559
& 4.084
& 3.284
& 4.058
& 0.947
& 0.105
& 0.972
& 0.829 \\

\midrule

\multicolumn{12}{l}{
\textit{(b) Controlled comparison of downstream TSE modeling}
} \\

LauraTSE
& FunCodec
& G
& --
& 3.609
& 4.084
& 3.336
& 4.333
& 0.908
& 0.159
& 0.974
& 0.876 \\

LauraTSE
& DualCodec
& G
& No
& 3.427
& 3.914
& 3.105
& 3.852
& 0.786
& 0.388
& 0.928
& 0.778 \\

Ours-Coupled
& DualCodec
& D--G
& No
& 3.431
& 3.972
& 3.167
& 3.904
& 0.835
& 0.260
& 0.940
& 0.801 \\

\textbf{Ours-Specialized}
& DualCodec
& D--G
& \textbf{Yes}
& 3.549
& 4.093
& 3.260
& 4.052
& 0.902
& 0.118
& 0.960
& 0.837 \\

\bottomrule
\end{tabular}
\caption{
What Makes Codec Tokens Effective for TSE: Disentanglement or
Specialization? ``G'' and ``D--G'' denote generative and
discriminative--generative modeling. ``Role'' indicates
whether semantic and acoustic representations are assigned task--aware modeling specialization roles.
}
\label{tab:representation_modeling}
\end{table*}

\begin{table}[t]
\centering
\small
\setlength{\tabcolsep}{5.0pt}
\begin{tabular}{lcccc}
\toprule
Variant
& NISQA$\uparrow$
& SBERT$\uparrow$
& dWER$\downarrow$
& WeSpk$\uparrow$ \\
\midrule
Full model
& {4.052}
& {0.902}
& {0.118}
& {0.837} \\

w/o sem condition
& 3.992
& 0.881
& 0.122
& 0.821 \\

w/o aco refiner
& 3.612
& 0.767
& 0.495
& 0.772 \\
\bottomrule
\end{tabular}
\caption{Ablation of semantic guidance and acoustic refinement.}
\label{tab:component_ablation}
\end{table}

\paragraph{Comparison with Existing TSE Systems.}
Since the generative baselines in Table~\ref{tab:main_results}
use different codec representations, we first compare models
under the same DualCodec. Ours-Specialized (task-aware framework) reduces the dWER
of Ours-Unified (LauraTSE framework)  from 0.388 to 0.118, while improving NISQA,
SpeechBERTScore, WavLM, and WeSpeaker from 3.852, 0.786,
0.928, and 0.778 to 4.052, 0.902, 0.960, and 0.837,
respectively. This controlled comparison confirms that the
gain arises from task-aware semantic--acoustic specialization
rather than codec replacement.

Our method also achieves lower dWER than most discriminative
baselines, including X-TF-GridNet, USEF-SepFormer, SpEx+,
and WeSep, while maintaining competitive perceptual quality.
The stronger USEF-TFGridNet-L still performs better in content
and speaker metrics. Compared with the generative baselines
with available dWER results, our method achieves the lowest
dWER of 0.118, outperforming LauraTSE, GenTSE, Metis, TSELM-L,
and LLaSE-G1, and provides a stronger balance
between content reliability, perceptual quality, and speaker
consistency. Such cross-system comparisons are partly limited
by the underlying codecs: DualCodec has a weaker reconstruction
upper bound than FunCodec on several task-relevant metrics.
Nevertheless, our method substantially improves the unified
DualCodec baseline.

\subsection{Ablation Study}
\paragraph{Representation Disentanglement Is Not Sufficient.}
Table~\ref{tab:representation_modeling} examines the interaction
between codec representation and modeling strategy. Panel~(a)
compares the codec reconstruction upper bounds. Although DualCodec
achieves higher DNSMOS scores, it underperforms FunCodec on
content- and speaker-related metrics, with a higher dWER
(0.105 vs.\ 0.080) and lower NISQA, SpeechBERTScore, WavLM, and
WeSpeaker scores. This indicates that semantic--acoustic
disentanglement introduces a representation trade-off rather than
a uniformly stronger codec.
Panel~(b) further shows that replacing FunCodec with DualCodec
under unified generative modeling degrades all downstream metrics:
dWER increases from 0.159 to 0.388, while NISQA and
SpeechBERTScore decrease from 4.333 to 3.852 and from 0.908 to
0.786, respectively. Thus, representation-level disentanglement
alone is insufficient when the semantic and acoustic tokens remain
coupled within a unified generative model.

\paragraph{Discriminative--Generative Coupling Modeing Is Also Insufficient.}
Adding a discriminative front-end partially alleviates the degradation of unified generative modeling, reducing dWER from 0.388 to 0.260 and improving SpeechBERTScore from 0.786 to 0.835. However, semantic extraction and acoustic reconstruction remain coupled rather than explicitly assigned to specialized pathways, limiting the benefit of semantic--acoustic disentanglement.

\paragraph{Task-Aware Modeling Specialization.}
Under the same DualCodec representation, Ours-Specialized
outperforms both LauraTSE-DualCodec and Ours-Coupled across
all reported metrics. Compared with the coupled discriminative--generative baseline, it
reduces dWER from 0.260 to 0.118, a relative reduction of 54.6\%,
while improving all other reported metrics. It further achieves a
69.6\% relative dWER reduction over unified generative modeling of
disentangled tokens, confirming that the gains arise from task-aligned
semantic and acoustic specialization rather than codec replacement or
generic discriminative--generative coupling. Although it does not
uniformly surpass FunCodec-based LauraTSE on perceptual and speaker
metrics, it reduces dWER from 0.159 to 0.118 despite the weaker
DualCodec reconstruction upper bound on several task-relevant metrics.

Overall, despite DualCodec's weaker reconstruction upper bound on
several task-relevant metrics, task-aware specialization substantially
improves its downstream TSE performance. This confirms our central
finding that effective codec-token TSE requires aligning representation
structure with appropriate modeling strategies, rather than relying on
semantic--acoustic disentanglement alone.

\paragraph{Roles of Semantic Guidance and Acoustic Refinement.}
Table~\ref{tab:component_ablation} shows that removing the semantic
condition increases dWER from 0.118 to 0.122 and degrades the other
metrics, indicating that semantic guidance supports target-content
preservation and interference suppression. Removing the acoustic
refiner causes a much larger degradation, increasing dWER to 0.495,
which confirms that coarse acoustic tokens alone are insufficient for
complete target-speech reconstruction. These results demonstrate the
complementary roles of semantic guidance and acoustic refinement.

\begin{table}[t]
\centering
\small
\setlength{\tabcolsep}{4.5pt}
\begin{tabular}{l l c c c c}
\toprule
Training
& Decoding
& OVRL $\uparrow$
& NISQA $\uparrow$
& dWER $\downarrow$
& WeSpk $\uparrow$ \\
\midrule
Split & AR  & 3.225 & 4.055 & 0.119 & 0.834 \\
Split & NAR & 3.256 & 4.091 & 0.113 & 0.848 \\
Joint & AR  & 3.260 & 4.052 & 0.118 & 0.837 \\
Joint & NAR & 3.258 & 4.092 & 0.097 & 0.844 \\
\bottomrule
\end{tabular}
\caption{Analysis of training and decoding strategies.}
\label{tab:training_decoding}
\end{table}

\paragraph{Training and Inference Strategies.}
In the standard setting, the decoder-only language model
generates target acoustic tokens sequentially, conditioned on
previously predicted tokens. Alternatively, NAR inference uses
the discriminative  output as pseudo acoustic labels,
avoiding recursive autoregressive prediction. Table~\ref{tab:training_decoding}
compares split and joint training under these two strategies.
Under AR decoding, joint training provides only marginal gains
over split training: DNSMOS-OVRL increases from 3.225 to 3.260,
while dWER decreases from 0.119 to 0.118. This suggests that
the pretrained semantic branch already provides an effective
target-related condition, and that the main gain arises from
task-aware specialization rather than joint optimization.NAR inference yields larger gains, reducing dWER to 0.113 under split training and 0.097 under joint training, while
maintaining comparable perceptual quality. These results suggest that NAR decoding can serve as a complementary inference strategy for improving content reliability.

\section{Conclusion and Future Work}
This paper investigates codec-token-based TSE from a semantic–acoustic disentanglement perspective. We show that representation disentanglement alone is insufficient and propose a task-aware discriminative–generative framework that specializes semantic extraction and acoustic reconstruction. Experiments on Libri2Mix demonstrate improved content reliability while maintaining competitive perceptual quality and speaker consistency. Future work will evaluate the framework with stronger dual-stream codecs and report the resulting findings to the community.

\bigskip

\bibliography{aaai2027}


\end{document}